\documentstyle[aps,twocolumn,epsfig]{revtex}

\newcommand{\beq}{\begin{equation}}
\newcommand{\eeq}{\end{equation}}
\newcommand{\bea}{\begin{eqnarray}}
\newcommand{\eea}{\end{eqnarray}}
\newcommand{\ba}{\begin{array}}
\newcommand{\ea}{\end{array}}
\newcommand{\bc}{\begin{center}}
\newcommand{\ec}{\end{center}}

\newcommand{\gsimeq}{\stackrel{>}{\scriptstyle\sim}}

\newcommand{\etal}{{\it et al.}}
\newcommand{\bml}{\begin{mathletters}}
\newcommand{\eml}{\end{mathletters}}

\newcommand{\commentout}[1]{{}}
\newcommand{\half}{\hbox{$1\over2$}}

\newcommand{\phidag}{\phi^\dagger}

\newcommand{\r}{{\bf r}}

\begin{document}
\draft
\wideabs
{
\title{Feshbach-Stimulated Photoproduction of a Stable Molecular
Condensate}

\author{Matt Mackie}
\address{Helsinki Institute of Physics, PL 64, FIN-00014
Helsingin yliopisto, Finland}
\date{\today}

\maketitle

\begin{abstract}

Photoassociation and the Feshbach resonance are, in 
principle, feasible means for creating a molecular 
Bose-Einstein condensate from an already-quantum-degenerate 
gas of atoms; however, mean-field shifts and irreversible decay
place practical constraints on the efficient delivery of
stable molecules using either mechanism alone. We therefore propose
Feshbach-stimulated Raman photoproduction, i.e., a combination of magnetic
and optical methods, as a means to collectively convert degenerate
atoms into a stable molecular condensate with near-unit efficiency.

\end{abstract}

}

\narrowtext

\section{Introduction}

In addition to translation, the molecular degrees of freedom include
rotation and vibration, so that the laser cooling techniques used to
create an atomic Bose-Einstein condensate (BEC)~\cite{BEC} are difficult,
if not impossible, to apply to molecules. Buffer gas
cooling~\cite{WEI98} and Stark-deceleration~\cite{BET00} methods
are thus being developed as replacements for lasers in the
precooling step towards a molecular BEC. Meanwhile, for systems where
atomic BEC is already available,
photoassociation~\cite{DRU98,JAV99} and the Feshbach
resonance~\cite{TIM99} have been proposed as a shortcut to the
production of a molecular condensate.

Photoassociation occurs when a pair of colliding ultracold atoms absorb a
laser photon~\cite{WEI99}, thereby jumping from the two-atom continuum to
a bound molecular state. If the initial atoms are
Bose-condensed, then the subsequent molecules
will also form a Bose-Einstein condensate~\cite{DRU98,JAV99}.
Nevertheless, photoassociation generally occurs to an excited
state, and the subsequent irreversible losses defeat the purpose of
molecular coherence. Adding a second
laser to drive molecular population to a stable electronic state,
stimulated Raman adiabatic passage (STIRAP)~\cite{BER98} in
photoassociation of a BEC~\cite{MAC00} has been proposed as a
means for avoiding radiative decay. However, collisions between
particles make it difficult to achieve adiabaticity, which ultimately
limits the practical efficiency of STIRAP to about fifty
percent~\cite{DRU01}.

On the other hand, the process known as the Feshbach
resonance~\cite{FES92} occurs when two ultracold atoms collide in the
presence of a magnetic field, whereby the spin of one of the
colliding atoms flips, and the pair jump from the two-atom continuum to a
quasibound molecular state. As with photoassociation, the so-formed
molecules will constitute a BEC if the incident atoms are themselves
Bose-condensed~\cite{TIM99,MBEC_FR}. Unfortunately, the quasibound
condensate, while translationally and rotationally ultracold, is
vibrationally very hot, and thus highly susceptible to collision-induced
vibrational relaxation. Hence, while an adiabatic sweep of the
magnetic field across the Feshbach resonance will create a
molecular BEC with near-unit efficiency, coherence is again moot in the
face of irreversible losses~\cite{MIE00}. Although adding a pair of lasers
to this scheme allows for a stable molecular condensate, the practical
efficiency of STIRAP is limited to about twenty percent~\cite{KOK01}.

The purpose of this Article is to develop a means for creating a stable
molecular BEC that is near-unit efficient in both
principle and practice. We therefore consider Feshbach-stimulated
photoproduction, wherein the quasibound state formed in the
presence of a static magnetic field is coupled via a Raman laser
configuration to a stable molecular state. Again, a version of this
scheme, in which STIRAP-inducing lasers are run concurrently with a sweep
of the magnetic field across the Feshbach resonance, has been previously
proposed~\cite{KOK01}; but, the efficiency of STIRAP is limited by the
time-dependence of the quasibound energy arising from the changing
magnetic field, and large system densities
($\gsimeq10^{14}$cm$^{-3}$) allow collision-induced mean-field shifts and
vibrational relaxation to dominate atom-molecule conversion. On the
contrary, we consider a static magnetic field interacting with a
modest-density gas of atoms, which minimizes the role of mean-field
shifts and vibrational relaxation, and enables the production of a
stable molecular condensate with near-unit efficiency.

The development herein is outlined as follows. Section~\ref{MODEL}
introduces the model, along with the concept of collective enhancement.
Section~\ref{NORAD} illustrates the creation of a stable molecular
condensate using a CW two-photon Raman configuration, as well as with
transient STIRAP, both of which avoid radiative losses. In
Sec.~\ref{QUENCHetc} we discuss vibrational quenching and
dissociation of the quasibounds, and estimate explicit numbers
to determine the atomic systems for which such losses are negligible.
Besides a summary, Section~\ref{SUM} considers our results in light of
recent experiments with $^{85}$Rb.

\section{Model}
\label{MODEL}

\begin{figure}
\centering
\epsfig{file=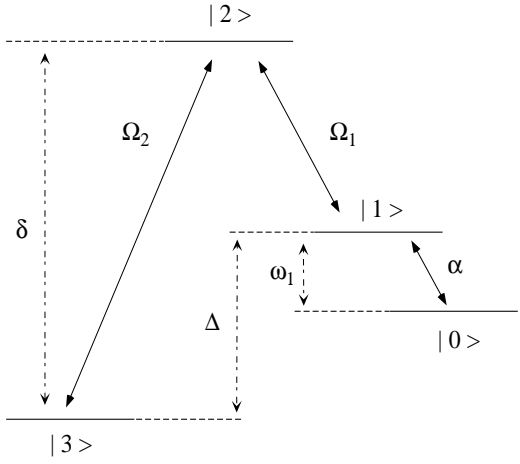,width=7cm,height=6cm}
\vspace{0.5cm}
\caption{
Few-level diagram corresponding to Feshbach-stimulated photoproduction of
a stable molecular condensate. For ease of illustration, we consider a
single-component gas of atoms that has Bose condensed into state
$|0\rangle$. The collective-enhanced Feshbach coupling is
$\alpha$, and the corresponding detuning is $\omega_1$. Similarly, the
laser 1 (2) coupling is $\Omega_{1(2)}$, and the two-photon (intermediate)
detuning is $\Delta$ ($\delta$). The full two-component case would involve
replacing the level $|0\rangle$ by a pair of (ideally) degenerate levels,
with a molecule formed by removing one atom from each level.}
\label{LEVELS}
\end{figure}

We model a two-component quantum degenerate gas of atoms coupled via a
Feshbach resonance to a condensate of quasibound molecules; additionally,
one laser drives quasibound $\leftrightarrow$ excited-bound molecular
transitions, thus creating a condensate of electronically excited
molecules, and a second laser induces excited-bound $\leftrightarrow$
ground transitions to a stable molecular condensate (see
Fig.~\ref{LEVELS}). The initial atoms are denoted by the field
$\phi_\pm(\r,t)$, the quasibound molecules by the field
$\psi_1(\r,t)$, the excited molecules by the field $\psi_2(\r,t)$, and,
finally, the ground state molecules by the field $\psi_3(\r,t)$. The
atoms may be either bosons or fermions, whereas the molecules are
strictly bosonic. All fields obey the appropriate commutation relations.
For the moment, we neglect both collisions and irreversible losses.
Assuming that these five modes sufficiently describe the
relevant physics, the Heisenberg equations of motion that govern
the evolution of the atomic and molecular fields are
\bml
\bea
i\dot\phi_\pm &=& (\mp)^\eta\,\half\bar\alpha\phidag_\mp\psi_1, \\
i\dot\psi_1 &=& \omega_1\psi_1 +\half\bar\alpha\phi_1\phi_2
  +\half\Omega_1\psi_2, \\
i\dot\psi_2 &=& \delta\psi_2
  +\half\Omega_1\psi_1 +\half\Omega_2\psi_3, \\
i\dot\psi_3 &=& \Delta\psi_3 +\half\Omega_2\psi_2.
\eea
\label{EQM}
\eml
Here $\eta=2$ (1) for bosons (fermions), and the amplitudes $\phi_\pm$
and $\psi_i$ have been taken as uniform over the volume of the system
($i=1,2,3$).

The strength of the Feshbach resonance, the process by which atoms
are converted into quasibound molecules, is given by
$\bar\alpha=[2\eta\pi |a|\mu_{ma}\Delta_R/m]^{1/2}$, where $a$ is the
off-resonance $s$-wave scattering length,
$\mu_{ma}$ is the difference in magnetic moments between the
quasibound molecule and free-atom pair, and $\Delta_R$ is the
resonance width in magnetic-field units. The
molecular energies are $\hbar\omega_i$, with the quasibound
energy is given specifically in terms of the magnetic field by
$\hbar\omega_1=\text{sgn}[a]\,(B-B_0)\mu_{ma}/2$, where
$B_0$ locates the resonance position. Note that the magnetic-field
dependence of $\bar\alpha$ and $\omega_1$ are chosen so that, upon
adiabatic elimination of the quasibound field, the resonance-induced
atom-atom scattering length has the correct dispersive form $\propto
\Delta_R/|B-B_0|$. Finally,
$\Omega_{1(2)}$ denotes the Rabi frequency corresponding to laser L1
(L2), which has a frequency $\omega_{L1(L2)}$; correspondingly, the
respective intermediate and two-photon detunings are
$\delta=\omega_2 -\omega_{L1}$ and $\Delta=\omega_3 -\omega_{L1}
+\omega_{L2}$.

We also scale the the atomic and molecular field
amplitudes in Eqs.~(\ref{EQM}) by the square-root of the total number
density of particles:
$x\rightarrow x'=x/\sqrt{\rho}$, with $x=\phi_\pm,\psi_1,\psi_2,\psi_3$.
Dropping the primes, the now order-unity field amplitudes evolve in time
according to
\bml
\bea
i\dot\phi_\pm &=& (\mp)^\eta\,\half\alpha\phidag_\mp\psi_1, \\
i\dot\psi_1 &=& \omega_1\psi_1 +\half\alpha\phi_1\phi_2
  +\half\Omega_{1}\psi_2, \\
i\dot\psi_2 &=& \delta\psi_2
  +\half\Omega_1\psi_1 +\half\Omega_2\psi_3, \\
i\dot\psi_3 &=& \Delta\psi_3
  +\half\Omega_2\psi_2.
\eea
\label{BE_EQM}
\eml
The term $\alpha=\sqrt{\rho}\,\bar\alpha$ would have previously
been referred to as the Bose-stimulated free-bound coupling~\cite{MAC00};
however, since the above equations of motion are statistics independent,
we realize that Bose stimulation (in this case) has nothing whatsoever to
do with bosons, but is instead a many-body effect that applies equally
well to Fermi-degenerate systems. This idea is implicit to a
\mbox{(Feshbach-)} photoassociation-induced Bose-Fermi
superfluid~\cite{TIM01}, as well as four-wave mixing~\cite{4WAVE} in a
Fermi-degenerate gas of atoms~\cite{DFG}. Hereafter, we therefore coin a
more appropriate term for
$\alpha$: the {\em collective-enhanced} free-bound coupling.

\section{Avoiding Radiative Decay}
\label{NORAD}

Given a Raman configuration, the brute-force way to avoid
radiative decay from an electronically-excited state is a large
intermediate detuning, and we thus consider the effective one-color
problem obtained by adiabatically eliminating the excited-bound molecular
field:
\bml
\bea
i\dot\phi_\pm &=& (\mp)^\eta\,\half\alpha\phidag_\mp\psi_1, \\
i\dot\psi_1 &=& \omega\,'_1\psi_1
  +\half\alpha\phi_1\phi_2 -\half\chi\psi_3, \\
i\dot\psi_3 &=& \Delta'\psi_3
  -\half\chi\psi_1,
\eea
\label{EFF_EQM}
\eml
where $\psi_1\leftrightarrow\psi_3$
transitions result from the two-color Rabi coupling
$\chi=\Omega_1\Omega_2/2\delta$, and the Feshbach (two-photon) detuning
has picked up a Stark shift
$\omega'_1=\omega_1-\Omega_1^2/4\delta$
($\Delta'=\Delta-\Omega_2^2/4\delta$). In such case, spontaneous
decay will now occur at a rate
$\propto\Gamma_s/\delta$, and is managed by choosing $\delta$
suitably large.

\begin{figure}
\centering
\epsfig{file=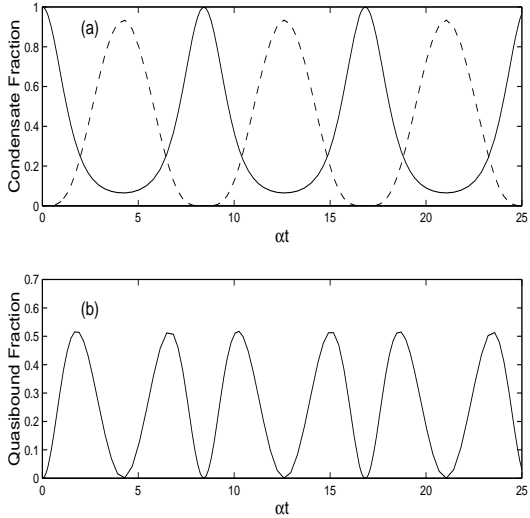,width=7cm,height=7cm}
\vspace{0.5cm}
\caption{
Feshbach-stimulated two-photon atom $\leftrightarrow$ stable molecule
conversion. Choosing
$\alpha=1$ sets the unit of frequency (time). The two-photon Rabi
frequency is $\chi=1$, and the system is tuned to Stark-shifted
resonance, i.e.,
$\omega'_1=\Delta'=0$. (a) The atomic (solid line) and stable
molecular (dashed line) condensates oscillate out of phase
as if there were a small yet nonzero detuning. (b) The detuning is
effectively induced by the presence of the quasibound condensate.}
\label{FESHPA}
\end{figure}

For simplicity, we focus on a
single-component BEC as our initial quantum gas: $\phi_\pm=\phi$,
$i\dot\phi=\alpha\phidag\psi_1$. The field
amplitudes are treated as
$c$ numbers, with initial conditions $\phi(0)=1$ and $\psi_i(0)=0$.
As shown in Fig.~\ref{FESHPA}, the atomic and stable molecular
condensates undergo Rabi-like
oscillations consistent with a nonzero but small detuning. This detuning
is effectively induced by the presence of the quasibound condensate, so
that, while the spontaneous decay of the excited-bounds is essentially
eliminated by a large intermediate detuning, losses due to vibrational
quenching and dissociation of the quasibounds would still be encountered.

Although irreversible decay from the electronically-excited molecules is
undoubtedly managed by a large intermediate detuning, we present a more
elegant alternative: stimulated Raman adiabatic passage.  The hallmark of
\mbox{STIRAP} is the {\em counterintuitive} pulse sequence~\cite{BER98},
which amounts to adjusting the two lasers so that initially, when most
everything is in the quasibound molecular condensate $\psi_1$, the
$\psi_3\leftrightarrow\psi_2$ coupling is strongest, and finally, when
effectively everything is in the stable molecular condensate $\psi_3$,
the $\psi_1\leftrightarrow\psi_2$ coupling is strongest. As the
population moves from quasibound to stable molecules, the state with the
larger population is always weakly coupled to the field $\psi_2$, thus
keeping the population of the excited molecular condensate low (ideally
zero) and reducing (eliminating) losses-- even for zero intermediate
detuning.

The idea is near that a properly-timed counterintuitive pulse sequence,
applied to a degenerate atomic gas in the presence of a static magnetic
field, will deliver a stable molecular BEC while avoiding radiative
losses. Given the appropriate frequency scale as the collective-enhanced
free-bound coupling
$\alpha$, then atoms will be coherently converted to quasibound
molecules on a timescale given by
$1/\alpha$, and we need only wait for sufficient
population to build before effecting
STIRAP. In our model we assume Gaussian laser pulses of the form
$\Omega_i=\Omega_0\exp[-(t-t_i)^2/T^2]$, where $\Omega_0$ determines the
pulse height~\cite{QB-EB_OK}, $t_i$ locates the pulse center, and $T$
defines the pulse width ($i=1,2$). 

\begin{figure}
\centering
\epsfig{file=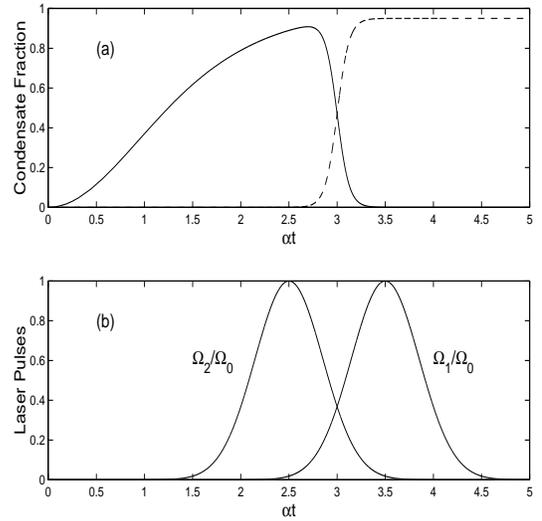,width=7cm,height=7cm}
\vspace{0.5cm}
\caption{
Creation of a stable molecular condensate via Feshbach-stimulated Raman
adiabatic passage. Frequency (time) is again in units
of $\alpha$. (a) As the collective conversion of atoms to quasibounds
nears completion (solid line), the lasers transfer the population to the
stable molecular condensate (dotted line) with ninety-five percent
efficiency. The maximum fraction of electronically-excited molecular
condensate (not shown) is of order $10^{-5}$. (b) The gaussian laser
pulses have equal widths and heights: $T=1/2$ and $\Omega_0=2\times10^3$.}
\label{FESHSTIRAP}
\end{figure}

This piece of intuition is correct, as shown in Fig.~\ref{FESHSTIRAP}.
We again use a single-component BEC as an example, with the pulse
parameters specified as $\alpha T=1/2$, $t_2=5\,T$, and $t_1=7\,T$.
Adiabaticity is assured by selecting pulse heights such that
$\Omega_0T=10^{3}$. The atoms are tuned close to the Feshbach resonance,
and the molecules onto laser resonance, by the choices
$\omega_1/\alpha=-10^{-2}$ and $\Delta=\delta=0$. The plot shows
the system coherently converting from atoms to quasibounds, and the laser
pulses subsequently transferring nearly the entire population to the stable
molecular condensate. Characteristic of a counterintuitive pulse sequence,
radiative losses are expectedly minimized by a low excited-state
population, of the order of
$10^{-5}$. Again, the presence of a meaningful fraction of quasibound
molecules means that the physics of quenching and dissociation must still
be addressed, to which we now turn.

\section{Quenching, Dissociation, and Explicit Numbers}
\label{QUENCHetc}

Given either a CW two-photon or transient STIRAP set-up, it is safe to
neglect radiative decay from the excited molecular state, and the
remaining practical issues involve the mean-field energy shift produced
by $s$-wave collisions between particles, as well as loss-inducing
vibrational relaxation and dissociation of the quasibound molecules.
Particle-particle collisions shift the system off of resonance by an
amount proportional to the $s$-wave scattering length; since the
numbers are not well known for atom-molecule and molecule-molecule
collisions, we take $\Lambda=4\pi\hbar\rho a/m$ as approximating the
mean-field shift due to $s$-wave particle collisions in general.
Meanwhile, vibrational relaxation in sodium occurs at a
rate~\cite{MBEC_FR} $2\Gamma\simeq\rho\,\times\,10^{-10}$~s$^{-1}$, and
we use this value to estimate the effects of both atom-molecule and
molecule-molecule vibrational quenching. Finally, dissociation of the
quasibounds back to the atomic continuum will occur analogous to
photodissociation~\cite{JAV99,KOS00,JAV02}, and necessarily involves
``rogue" modes defined as modes lying outside the atomic condensate
(Fermi sea). All told, rogue dissociation is {\it non-exponential}, i.e.,
the Fermi Golden Rule and simple inclusion with a non-hermitian
term~\cite{MBEC_FR,KOK01} do not strictly apply~\cite{JAV02}, and will
impose a rate limit on all-bosonic atom-molecule conversion given roughly
by $R_L=6\,\omega_\rho$, where
$\omega_\rho=\hbar\rho^{2/3}/m$~\cite{JAV99,KOS00,JAV02}. For
fermionic atoms, the factor of two difference in the collective-enhanced
free-bound coupling $\alpha$ should lead to a doubling of the maximum
rate, $R_L=12\,\omega_\rho$~\cite{ROGUE_FERMI}. From
Fig.~\ref{FESHSTIRAP}, ninety-five percent of the atoms are converted in
$3/\alpha$ seconds, corresponding to a rate
$R=0.95\,\alpha/3$.

Explicit numbers for various
systems~\cite{HUL02,INO98,RUB85,VAN01,6Li,40K} are listed in
Table~\ref{TABLE}, where the densities (resonance widths) were
chosen as a balance between being low (high) enough to enable safe
neglect of $s$-wave and vibrational-quench-inducing collisions, and large
(small) enough to keep atom-molecule conversion below the rate
limit.  The table indicates that rogue dissociation is the main obstacle.
Overall, the $^{23}$Na~(853~G) and $^{87}$Rb~(680~G) resonances appear as
the strongest candidates. For a CW two-photon Raman scheme,
intermediate detunings $\delta\simeq 10^3\:\mu$s$^{-1}$ should manage
typical spontaneous molecular decay rates $\Gamma_s\simeq
10\:\mu$s$^{-1}$ over the timescale $\alpha^{-1}$, while at the same
time allowing moderate bound-bound Rabi frequencies:
$\Omega_1=\Omega_2\simeq 0.1$~ms$^{-1}$ for $\chi/\alpha=1$. The
\mbox{STIRAP} peak Rabi frequency $\Omega_0\simeq 1\:\mu$s$^{-1}$ is
larger, but not unreasonable. If by chance irreversible quasibound decay
proves more than a marginal issue, one might imagine a multi-shot STIRAP
scheme (see also Ref.~\cite{KOK01}), where a smaller fraction of the
population is converted per shot, and losses are further reduced by
keeping the population low.

\section{Summary}
\label{SUM}

Admittedly, our inclusion of fermions is
based on a BEC as an explicit example; nevertheless, the model itself is
not dependent on statistics. Collective enhancement applies equally to
fermions and bosons, so that the atom-quasibound conversion
should still set the timescale $\sim 1/\alpha$, as is evident in
Ref.~\cite{TIM01}. Absent Pauli blocking, dominant rogue losses
render the production of stable molecules inefficient for the $^6$Li and
$^{40}$K resonances considered; nevertheless, in a manner analogous to the
bosonic case~\cite{JAV02} (see also below), collective oscillations
between atoms and correlated dissociation pairs, i.e., {\em Cooper
pairs}, could still occur.

We have, moreover, identified the 155~G $^{85}$Rb resonance as a severely
rogue-dominated system ($R/R_L\approx10$), meaning that it is useless
for producing any significant fraction of molecular condensate, stable or
otherwise. On the other hand, Feshbach-only experiments have led to the
observation of bursts of atoms emanating from a remnant
condensate~\cite{DON01}, and follow-up experiments using pulsed
magnetic fields indicate large-amplitude oscillations ($\sim$~30\%)
between remnant and burst atoms with a good many of the initial atoms
gone missing-- prompting speculation as to the formation of a molecular
condensate~\cite{DON02}. Although we have no doubt as to the achievement
of atom-molecule coherence, the amplitude of the experimental oscillations
need not be indicative of the number of molecules present: the
double-pulse experiments~\cite{DON02} realize a Ramsey-type~\cite{RAM50}
interferometer, which maps out collective oscillations
between atoms and dissociated atom pairs, and which is highly sensitive
to a small fraction of molecular condensate ($\sim$~1\%)~\cite{MAC02} (see
also Ref.~\cite{KOK02}).

In conclusion, we have presented a method for preparing a stable
molecular condensate with near-unit efficiency: (i) start with a quantum
degenerate gas of atoms, a magnetic field tuned at or near a Feshbach
resonance, and a pair of lasers arranged in a Raman
configuration; (ii) induce either CW two-photon Rabi flopping or transient
STIRAP; (iii) obtain a stable molecular condensate. Favorable systems for
implementation are $^{23}$Na and $^{87}$Rb. The advantage of this scheme
is near-unit efficiency, which implicitly avoids rather large
inelastic losses incurred by molecular collisions with unconverted
atoms~\cite{WYN00}, and delivers a truly stable condensate of molecules.
Moreover, experiments with photoassociation alone~\cite{WYN00} are just
on the verge of coherent conversion~\cite{KOS00}, and culmination requires
a difficult-to-impossible balancing of low density and high laser
intensity to simultaneously minimize mean-field shifts and rogue
dissociation; a magneto-optical scheme may therefore provide the best
opportunity for observing large-scale collective atom-molecule
oscillations.

\section{Acknowledgements}

The author thanks Randy Hulet, Juha Javanainen,
Emil Lundh, Jani
Martikainen, Jyrki Piilo, and Kalle-Antti Suominen for helpful
discussions, and acknowledges the Academy of Finland for support
(projects 43336 and 50314).

\begin{table}
\centering
\caption{Explicit numbers for various systems.
Resonance positions are given in~G, densities in
$10^{14}$cm$^{-3}$, and atom-molecule couplings in
ms$^{-1}$. To estimate
$\alpha$, we assume $\mu_{ma}=\mu_B$, where
$\mu_B$ is the Bohr magneton.}
\begin{tabular}{rcccccc}
Atom\hspace{0.4cm}($B_0$) & $\rho$ & $\alpha$
  & $\Lambda/\alpha$ & $\Gamma/\alpha$ & $R/R_L$ \\
\hline \vspace{-0.25cm}\\
$^7$Li~\cite{HUL02} (800) & 1 & 3989 & 0.0008 & 0.0013 & 1.077 \\
$^{23}$Na~\cite{INO98} (853) & 0.01 & 4.13 & 0.029 & 0.012 & 0.079 \\
$^{85}$Rb~\cite{RUB85} (155) & 0.01 & 142.4 & -0.0013 & 0.0004 & 10.1 \\
$^{87}$Rb~\cite{VAN01} (680) & 0.1 & 8 & 0.061 & 0.062 & 0.013 \\
$^6$Li~\cite{6Li} (800) & 0.1 & 9635 & 0.0007 & 0.00005 & 5.17 \\
$^{40}$K~\cite{40K} (190) & 1 & 1264 & 0.015 & 0.004 & 0.974 \\
\end{tabular}
\label{TABLE}
\end{table}


\begin{references}
\bibitem{QB-EB_OK}
Unlike the free $\leftrightarrow$ excited-bound
case~\cite{DRU01}, for a system tuned close to the Feshbach resonance,
the dipole matrix element for the quasibound $\leftrightarrow$
excited-bound transition should in general be sufficiently large to allow
for equal pulse heights. See, e.g., Ref.~\cite{KOK01}.
\bibitem{ROGUE_FERMI}
  Pauli blocking should not affect the rate limit due to the large number
of available dissociation-continuum states.
\end{references}
\end{document}